\documentclass[twocolumn,prl,preprintnumbers,amsmath,amssymb,superscriptaddress,unsortedaddress]{revtex4}

\usepackage{graphicx}
\usepackage{dcolumn}
\usepackage{bm}
\usepackage[Symbol]{upgreek}  

\begin{document}


\title{Superconducting Quantum Interference Device Amplifiers with over 27 GHz of Gain-Bandwidth Product Operated in the 4 GHz--8 GHz Frequency Range\footnote{This paper is a contribution of the U.S. government and is not subject to U.S. copyright.}}

\author{Lafe Spietz}
 \email{lafe@nist.gov}
\author{Kent Irwin}
\author{Jos\'e Aumentado}%
\affiliation{National Institute of Standards and Technology, Boulder, Colorado 80305, USA}

\date{\today}

\begin{abstract}
	We describe the performance of amplifiers in the 4 GHz--8 GHz range using Direct Current Superconducting Quantum Interference Devices(DC SQUIDs) in a lumped element configuration.  We have used external impedance transformers to couple power into and out of the DC SQUIDs.  By choosing appropriate values for coupling capacitors, resonator lengths and output component values, we have demonstrated useful gains in several frequency ranges with different bandwidths, showing over 27 GHz of power gain-bandwidth product.  In this work, we describe our design for the 4 GHz--8 GHz range and present data demonstrating gain, bandwidth, dynamic range, and drift characteristics.
\end{abstract}

\maketitle

	The development of low-noise microwave amplifiers is an increasingly critical need in low-temperature physics.  One technology that has proven to have some of the best characteristics in terms of gain-bandwidth product and overall noise performance is the DC SQUID(Direct Current Superconducting Quantum Interference Device)-based microwave amplifier\cite{mueck1,mueck_ghz,prokopenko1}.  The SQUID has been shown to have better noise than similar semiconductor-based amplifiers\cite{mueck_quantum}, and better gain-bandwidth product, ease of use and power handling capabilities than parametric amplifiers that operate at or below the standard quantum limit\cite{manuel_2008}.  The SQUID also has power dissipation as much as four orders of magnitude lower than HEMT(High Electron Mobility Transistor) amplifiers.  

	In recent years, the 4 GHz--8 GHz frequency range, known as the C-band, has proven to be of great importance for a wide variety of quantum measurement experiments\cite{cQED1}, and so we have chosen to focus our design efforts primarily on that band.  Based on the model of the input impedance of a lumped-element DC SQUID described and experimentally verified elsewhere\cite{spietz_squidamp1}, we have designed input transformers with hundreds of megahertz of instantaneous bandwidth at frequencies in the range of 4 GHz--8 GHz.  Our amplifiers consist of a quarter-wave resonator with a coupling capacitor in the 50 fF--100 fF range coupled to a section of transmission line varying in length from 1.4 mm to 3.5 mm, which feeds into the input coil of the SQUID.  The output of the SQUID is then coupled to the 50 $\Omega$ transmission line by a multipole transformer on the chip(see Fig. 1).  The purpose of these measurements was threefold.  Firstly, we wanted to show that we could build amplifiers in the 4 GHz--8 GHz range with useful gains and bandwidths.  Secondly, we wanted to show that we could tune the center frequencies of the amplifiers' gains by changing the length of the input resonator in a predictable way that would allow us to design for specific frequencies.  Finally, by measuring amplifiers with a variety of input resonator lengths we were able to characterize the imaginary component of the input impedance, which will be necessary for future, more complex input transformer designs.  
	
\begin{figure}
\includegraphics{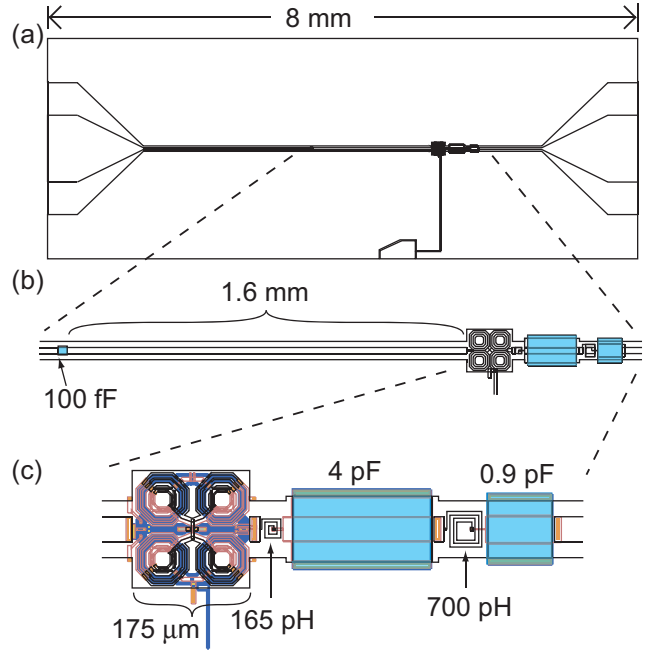}
\caption{Diagram of 100 fF amplifier chip with three levels of magnification (a,b and c), adapted from CAD files used for fabrication.  The output transformer consists of a set of two overlap capacitors to ground and two series spiral inductors, designed to match over as much of the target band as possible.  Another overlap capacitor couples the input line to the quarter wave resonator, which is terminated in the input coil of the SQUID.}
\end{figure}		
	
	Our SQUIDs, also described in reference \cite{spietz_squidamp1}, are designed to behave as lumped-element components by keeping the physical size of the SQUID below 200 $\upmu$m and use a slotted washer to minimize input stray capacitance, moving parasitic self-resonances to frequencies well above the measurement band.  The SQUIDs are also designed with a second-order gradiometer configuration to minimize magnetic pickup as well as the self inductance of the SQUID loop (18 pH)\cite{squidref1,squidref2,squidref3}.  All SQUIDs are fabricated by use of a Nb/AlOx/Nb optical trilayer process with 60 $\upmu$A critical current for each junction, shunted by 2.3 $\Omega$ resistors, and with two wiring layers and two insulating layers of SiO$_2$.  The flux-to-voltage transfer function $\partial V/\partial\phi$ at the flux and current bias points with useful gain was in the range of 200 $\upmu V/\Phi_0$--300 $\upmu V/\Phi_0$.  Each SQUID has a dc flux bias coil that winds half a turn around each of the four lobes for the dc flux bias, and a 600 pH microwave flux bias coil that winds a turn and a half around each lobe for the microwave input coil.

\begin{figure}
\includegraphics{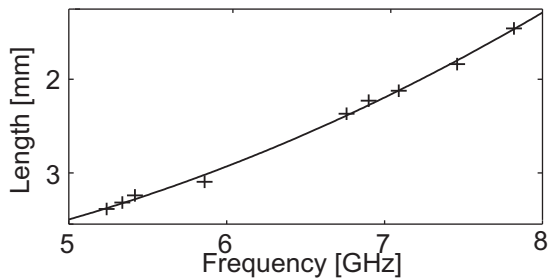}
\caption{Length of resonator as a function of frequency of best input match for a 60 fF input capacitor amplifier. Each point represents a set of measurements on a separate amplifier with a different length input resonator.  The continuous line is a fit to a second-order polynomial. Note that the frequency dependence is smooth, allowing for interpolation as needed to design for any frequency in the band.  Data are displayed with frequency as the independent variable since this is intended to be used as a design tool where the frequency is selected based on some application, and the length is chosen to match that frequency.}
\end{figure}

	We measured the input return loss of amplifiers with several different lengths in order to determine empirically how to design for different frequencies in the 4 GHz--8 GHz band and in order to determine the imaginary component of the input impedance as a function of frequency.  For each length of resonator, we measured the returns loss over a range of flux bias and current bias.  For each return loss curve, we extracted the frequency at which the most power was absorbed, and then put that frequency into a weighted average frequency, weighting for the depth of the resonance in linear power units.  By taking such weighted averages, we were able to compare a large quantity of data on each amplifier by use of a single average frequency number.  Fig. 2 plots the resonator length for a given center frequency against this weighted average matching frequency.  An understanding of the quarter wave input circuit permits us to use these data to find that the imaginary component of the input impedance ranges from about 30 $\Omega$ to about 60 $\Omega$.  We used similar methods to determine the real component of the input impedance, which we presented in an earlier paper \cite{spietz_squidamp1}.

	In addition to the reflection measurements described above, we measure the power gain of our amplifiers at the base temperature(approximately 40 mK) of a dilution refrigerator.  We use a cold microwave transfer switch, to bypass the amplifier and obtain an \textit{in situ} calibration.  Thus, our gain measurement determines the usable power gain from the input SMA connector of the amplifier box to the output SMA connector\cite{agilent1}.  
		
\begin{figure}
\includegraphics{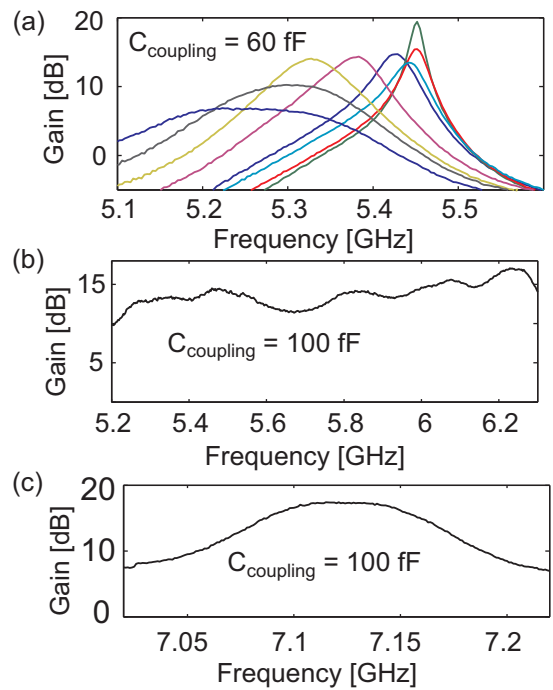}
\caption{(a) Gains for several flux bias points, ranging over about 5 \% of a flux quantum, and fixed 150 $\upmu$A current bias of the amplifier with a 60 fF coupling capacitor.  Note the trade-off between gain and bandwidth in the different curves.  (b) High bandwidth bias point(current bias of approximately 140 $\upmu$A) for 100 fF coupling capacitor amplifier.  The data shown on this plot demonstrate over 27 GHz of gain-bandwidth product (the integral of linear power gain over the frequency range).  (c)  Same bias point and amplifier as (b), showing that there are multiple gain maxima, and that feedback effects cause surprisingly complex frequency dependence in the gain.}
\end{figure}
		
	Fig. 3 shows several gain curves for two different amplifiers at different flux and current bias points.  The frequency dependence of the gain is a complicated function of the bias points.  We find that for amplifiers with a 60 fF input coupling capacitor, the gain curves are smooth and simple, showing the trade-off between gain and bandwidth.  While these gains and bandwidths are enough to be useful for many applications, more bandwidth is desirable, and we find that the bandwidths of the 100 fF input capacitor amplifiers are much higher.  

	In both the gain data and the return loss data we observe frequencies and bandwidths that differ widely from those expected from simply the frequency dependence of the input and output transformers and the SQUID.  If a resonator coupled to a gain element such as a SQUID has some feedback mechanism, the magnitude and phase of that feedback can have profound effects on the frequency dependence of the overall circuit, leading to drastic suppression or enhancement of Q as well as shifts in frequency by many line widths.  Although one might expect only moderate increases in bandwidth going from 60 fF to 100 fF, we observe increased bandwidth by as much as an order of magnitude, with over 1 GHz of potentially useful gain for some bias points in the 100 fF amplifier.  We believe that this is due to these complex feedback effects of the SQUID, and that greatly increased bandwidth should be possible with further study of the SQUIDs intrinsic S-parameters similar to that carried out with microwave transistors.

\begin{figure}
\includegraphics{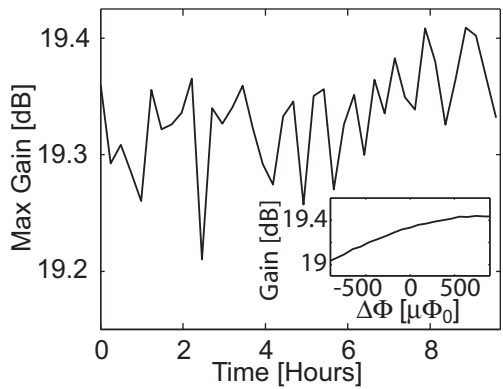}
\caption{(a) Total microwave measurement chain drift as a function of time overnight.  Inset (b) shows dependence of maximum gain on the flux bias.}
\end{figure}

	We find, via a set of drift and bias dependence measurements, that the amplifier gain is stable over many hours.  Fig. 4 shows both the way that gain depends on flux bias in a small range of biases, and how the long-term stability of the amplifier corresponds to possible flux drift.  The drift data were taken by recording a series of gain traces as a function of frequency at a range of flux bias points every fifteen minutes for approximately 10 hours.  Between measurements, both the flux and current bias were turned off.  Thus they show not just the stability of the amplifier, but the repeatability of the bias.  This repeatability is important for a practical amplifier, because that makes it possible to use the same bias point each time.  It is worth noting that this drift plot represents the entire microwave measurement chain, including two semiconductor amplifiers (one at 4 K and one at room temperature) and several cables and passive components, all of which could contribute to the observed drift.  These drift data also show the value of the magnetic noise immunity provided by the second-order gradiometer configuration of the SQUIDs.  We had one layer of high permeability magnetic shielding, and no superconducting shield.  Movement of large ferrous objects near the dewar had no observable effect on the bias point of the SQUID.  
	
\begin{figure}
\includegraphics{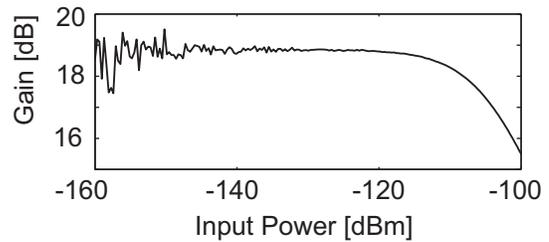}
\caption{Gain at fixed bias point as a function of amplifier input power.}
\end{figure}	
	
	Another important figure of merit in any amplifier is its power-handling capability, which determines the dynamic range.  We have measured the gain as a function of input power for a variety of bias points, and find that while the power-handling capabilities fall short of HEMT amplifiers, they are high enough to be useful.  The 1 dB compression point is at approximately -110 dBm, which should be more than sufficient for most applications where a near-quantum-limited amplifier is required.
	
	In conclusion, this amplifier has the instantaneous bandwidth, center frequencies, power handling and stability required to be useful for a range of quantum-measurement experiments. We are still measuring noise performance, but preliminary SNR improvement measurements show that system noise can be improved by an order of magnitude over typical measurements with HEMT amplifiers.  With careful selection of bias point and input circuit we believe this amplifier could, in its present form, lead to improvements in dispersive qubit readouts\cite{rob_private}.   The gain-bandwidth product demonstrated implies that it should be possible to construct an amplifier with a full 4 GHz--8 GHz range by sacrificing gain for bandwidth, then cascading two or three amplifiers together.  An amplifier with these characteristics would greatly benefit the low-temperature physics community.
	
\begin{acknowledgments}
We thank Konrad Lehnert, Michel Devoret, Dan Schmidt, Michael Elsbury, and Rob Schoelkopf for useful discussions.  We thank NSA for support.
\end{acknowledgments}

\end{document}